\documentclass[journal]{new-aiaa}

\usepackage{amssymb}

\usepackage{amsmath,amsfonts}
\usepackage{multirow}
\usepackage{hyperref}
\usepackage{lineno}
\usepackage{hyperref}
\usepackage{graphicx}
\usepackage{subcaption}
\usepackage{bm}
\usepackage{dcolumn}
\usepackage{booktabs}
\usepackage{lineno}
\usepackage{empheq}
\usepackage{mathtools}
\usepackage{booktabs}
\usepackage{makecell}
\usepackage{cellspace}
\usepackage{caption}
\usepackage{xcolor}
\usepackage{scalerel,stackengine}
\usepackage{xspace}
\usepackage[version=4]{mhchem}
\usepackage{siunitx}
\usepackage{tikz,pgfplots,pgfplotstable}
\usepackage{subcaption}
\usepackage[mathscr]{euscript}

\DeclareMathAlphabet{\mathcalligra}{T1}{calligra}{m}{n}
\usepackage{setspace}

\newcommand\sgs{\mathrm{sgs}}                

\makeatletter
\def\ps@pprintTitle{%
\let\@oddhead\@empty
\let\@evenhead\@empty
\def\@oddfoot{\reset@font\hfil\thepage\hfil}
\let\@evenfoot\@oddfoot
}
\makeatother

\title{Numerical Analysis of Pure and Blended Fuel Sonic Jets in a Mach 2 Crossflow}

\author{Radouan Boukharfane\footnote{radouan.boukharfane@um6p.ma, AIAA Member}}
\affil{Mohammed VI Polytechnic University (UM6P), College of Computing, Benguerir, Morocco}

\begin{document}
\date\today
\maketitle

\begin{abstract}
The injection of transverse jets into supersonic compressible crossflows represents a fundamental configuration relevant to a spectrum of high-speed applications. The intricate interactions arising between the crossflow and the injected jet induce complex flow phenomena, including shock waves and vortical structures, the characteristics of which are significantly contingent upon the thermophysical properties of the injected fuel. While prior investigations have addressed the influence of various fuels, a knowledge gap persists concerning the behaviour of alternative and synthetic multicomponent fuels within this flow regime. The present work employs high-fidelity large-eddy simulations (LES) to examine the impact of ten distinct fuels—hydrogen, methane, ethylene, ammonia, syngas mixture, and a synthetic blend, alongside several \ce{NH3}/\ce{H2}/\ce{N2} mixtures—on the macroscopic flow structures and mixing attributes within a transverse sonic jet immersed in a Mach 2 crossflow. By maintaining a uniform momentum flux ratio across the investigated cases, the study aims to isolate the influence of the unique thermophysical properties of each fuel on the windward mixing layer, a region critically important for initial entrainment processes. The investigation quantifies the effects of molecular weight, heat capacity ratio, and density on the development and evolution of coherent structures through a detailed examination of instantaneous flow fields, vortex dynamics, scalar distributions, and turbulence statistics. The results are expected to provide pertinent insights into fuel-dependent mixing mechanisms in supersonic flows, thereby contributing to the advancement of more efficient and versatile propulsion systems.
\end{abstract}

\section{Introduction}

Transverse jets in supersonic compressible flows (JISCF) are an archetypal setup common to numerous various engineering and scientific problems. It encompasses, but is not restricted to, air-breathing hypersonic propulsion vehicles (e.g., scramjets), active flow control devices, chemical plume dispersion, fuel injection methods for ram/scramjet flight, and environmental risk reduction strategies with regard to dispersion of high-speed jets. Physical processes governing JISCF are rich due to multiscale interaction among the injected jet and the high-speed crossflow, such as shear layers, shock structures, vortical instability, and strong compressibility effects \cite{mahesh2013interaction}. Fundamental understanding of fluid dynamics and mixing phenomena in such configurations is therefore relevant to advancing propulsion, high-speed aerodynamics, and reactive flow systems technologies. One of the most popular and most researched injection techniques is the transverse sonic injection through a circular orifice. In this setup, a sonic jet is injected perpendicularly into a supersonic freestream, thereby producing a highly complex flow field. The interaction produces a bow shock in the upstream region of the jet, a barrel shock inside the jet core, and strong shock–boundary layer interactions. Most importantly, the jet-crossflow interaction generates large vortical structures, in particular a counter-rotating vortex pair (CVP), which is accountable for freestream air entrainment into the jet and the ensuing mixing process \cite{gruber1995mixing}. The behavior and character of these vortical structures rely heavily on the characteristics of the injected fuel.

Numerous experimental studies have reported significant differences in jet penetration, structural deformation, and mixing efficiency between fuels with different molecular weights and specific heat ratios. For example,
\citet{gruber1997compressibility} and \citet{gruber1995mixing} used planar laser-induced fluorescence (PLIF) in combination with Rayleigh scattering methods to investigate the mixing characteristics of various light fuels, such as hydrogen and helium. Less dense fuel was discovered to create larger and more organized vortical structures, which increase fuel-air mixing because of greater contact with surrounding flow. Conversely, heavier fuels like ethylene or methane are observed to induce slower convection of the jet and lower penetration, as illustrated by \citet{ben2006time}, leading to various mechanisms of mixing layer formation.
Complementary numerical studies based on large-eddy simulation (LES) and direct numerical simulation (DNS) have highlighted the importance of compressibility, the heat capacity ratio, and density contrast in modifying turbulent transport dynamics. For instance,
\citet{kawai2010large} and \citet{watanabe2012large} demonstrated that the growth of shear-layer instabilities and the evolution of jet morphology are strongly influenced by the thermodynamic properties of the fuel, even when the momentum flux ratio is held fixed. Besides, \citet{won2010numerical} highlighted that coherent structures developing from the windward shear layer in the vicinity of the jet orifice are accountable for entrainment of ambient fluid, and that these structures are extremely sensitive to the influence of fuel compressibility.

Although great advances have been made in this area, accurate prediction of turbulent mixing and scalar transport is still an issue. Achieving this objective not only demands careful selection of turbulence models but also a suitable grid resolution. Furthermore, besides these endeavors, the effects of alternative and synthetic fuels on supersonic mixing phenomena are also not well understood. This mismatch is especially significant in the case of multicomponent fuels, such as syngas and ammonia mixtures, that are being progressively contemplated for their application in low-carbon propulsion systems. These fuels display intermittent thermochemical characteristics, such as lower or variable specific heat ratios and densities, all of which have a profound effect on jet dynamics and the production of turbulence. The present research intends to bridge this gap by examining the influence of fuel composition on the macroscopic structures and mixing characteristics in a transverse sonic jet injected into a Mach 2 supersonic crossflow. To do so, high-fidelity LES are used to investigate six fuels: hydrogen (\ce{H2}), methane (\ce{CH4}), ethylene (\ce{C2H4}), ammonia (\ce{NH3}), a syngas mixture (\ce{CO} + \ce{H2}), and a synthetic mixture of \ce{NH3}, \ce{H2}, \ce{O2}, and \ce{N2}. The simulation setups for all cases are configured to have the same momentum flux ratio between the jet and the crossflow, allowing fuel-specific thermophysical property effects to be examined in isolation. Of specific concern in this study is the windward mixing layer, which is a region of paramount significance for entrainment and initial mixing. Through comparative measurements of instantaneous flow fields, vortex dynamics, scalar concentration distribution, and turbulence statistics of various fuels, this study endeavors to quantify the influence of fuel-dependent physical properties—namely, molecular weight, heat capacity ratio, and density—on the development and evolution of coherent structures. The results of this study will provide valuable insights on the fuel-dependent mixing regime in supersonic flows, with applications to the design of more efficient and fuel-flexible propulsion systems.

\section{Methodology}
\label{sec:Methodology}
\subsection{Filtered compressible Navier--Stokes equations}
%

The spatial domain is defined by a Cartesian coordinate system where $x_1$ (or $x$), $x_2$ (or $y$), and $x_3$ (or $z$) are the streamwise, wall-normal, and spanwise directions, respectively. The dynamics of the turbulent flow is simulated using Large Eddy Simulation (LES) based on the Favre-filtered (density-weighted) compressible Navier-Stokes equations. The Favre filtering operation on a general field variable $f$ is defined as $\widetilde{f} = \overline{\rho f} / \overline{\rho}$, where $\rho$ is the instantaneous density and the overbar indicates the spatial filtering operation:
$$
\overline{f}(\boldsymbol{x},t) = \int G\left(\boldsymbol{x} - \boldsymbol{x'}\right)f(\boldsymbol{x'},t)\,\mathrm{d}\boldsymbol{x'} \,,
$$
with $G\left(\boldsymbol{x}\right)$ being the filter kernel \cite{leonard1975energy}. Following this filtering procedure, the conservation equations for mass, momentum, and species mass fractions are given by:
\begin{equation}\label{eq-masse_filtre_sci}
\partial_t \overline{\rho}+\partial_j\left(\overline{\rho} \widetilde{u}_j\right)=0 \,,
\end{equation}
\begin{equation}\label{eq-qdm_filtre_sci}
\partial_t\left(\overline{\rho} \widetilde{u}_i\right)
+\partial_j\left(\overline{\rho} \widetilde{u}_i \widetilde{u}_j\right)=
-\partial_i \overline{p}
+\partial_j\overline{\tau}_{ij}
-\partial_j \left(\overline{\rho u_i u_j}-\overline{\rho}\widetilde{u}_i\widetilde{u}_j\right) \,,
\end{equation}
\begin{equation}\label{eq-fracmass_filtre_sci}
\partial_t\left(\overline{\rho} \widetilde{Y}_\alpha\right)
+\partial_i\left(\overline{\rho} \widetilde{u}_i \widetilde{Y}_\alpha\right)=
-\partial_i\left(\overline{\rho Y_\alpha {V}_{\alpha i}}\right)
-\partial_i\left(\overline{\rho Y_\alpha u_i}-\overline{\rho} \widetilde{Y}_\alpha \widetilde{u}_i\right)
+\overline{\rho}\widetilde{\dot\omega}_\alpha \,,
\end{equation}
where $t$ is the temporal coordinate, and $\partial_t = \partial/\partial t$ and $\partial_i = \partial/\partial x_i$ represent the partial derivative operators. The Einstein summation convention is implied for repeated indices. The filtered density is denoted by $\overline{\rho}$, the filtered pressure by $\overline{p}$, the Favre-filtered velocity vector by $\widetilde{\boldsymbol{u}}$, and the filtered mass fraction of the $\alpha$-th species by $\widetilde{Y}_{\alpha}$ ($\alpha \in \{1, \dots, \mathcal{N}_{\text{sp}}\}$, with $\mathcal{N}_{\text{sp}}$ being the total number of chemical species). These filtered thermodynamic variables are related through the ideal gas equation of state:
\begin{equation*}
\overline{p} = \overline{\rho} \mathsf{R} \widetilde{T}/\widetilde{\mathcal{W}},
\end{equation*}
where $\widetilde{\mathcal{W}}^{-1} = \sum_{\alpha=1}^{\mathcal{N}_{\text{sp}}} \widetilde{Y}_\alpha/\mathcal{W}_\alpha$, $\mathcal{W}_\alpha$ is the molecular weight of species $\alpha$, and $\mathsf{R}$ is the mixture’s gas constant. The terms ${V}_{\alpha i}$ and $\dot\omega_\alpha$ represent the diffusion velocity and the chemical production rate of species $\alpha$, respectively. The resolved viscous stress tensor is modeled as:
\begin{equation*}
\overline{\tau}_{ij} = 2\mu(\widetilde{T}) \left(\widetilde{S}_{ij}-\frac{1}{3}\widetilde{S}_{kk} \delta_{ij}\right) \,,
\end{equation*}
where $\widetilde{S}_{ij} = (\partial_j\widetilde{u}_i+\partial_i\widetilde{u}_j)/2$ is the resolved strain-rate tensor.
In this study, the molecular diffusion fluxes are modeled using a mixture-averaged approach, based on a modified Hirschfelder and Curtiss approximation \cite{hirschfelder_molecular_1964}. Furthermore, it is assumed that the filtered molecular diffusion flux of species $\alpha$ can be expressed analogously to its instantaneous counterpart, applied to the filtered variables:
\begin{equation}\label{eq-diff-mix_sci}
\overline{\rho Y_\alpha V_{\alpha i}}\approx
\overline{\rho} \widetilde{Y}_\alpha \widetilde{V}_{\alpha i}=
-\overline{\rho} \widetilde{D}^m_\alpha\mathcal{W}_\alpha\partial_i\widetilde{X}_{\alpha}/\widetilde{\mathcal{W}}
+\overline{\rho} \widetilde{Y}_\alpha \widetilde{V}_i^c \,,
\end{equation}
where $\widetilde{D}^m_{\alpha}$ is the resolved mixture-averaged diffusion coefficient for species $\alpha$. The last term in Eq.~\eqref{eq-diff-mix_sci} represents a correction velocity, $\widetilde{V}_i^c = \sum_{\beta=1}^{\mathcal{N}_{\text{sp}}} \widetilde{D}^m_\beta \left( \mathcal{W}_\beta / \widetilde{\mathcal{W}} \right) \partial_i \widetilde{X}_\beta$, introduced to ensure global mass conservation at each time step by maintaining consistency between the discrete species mass fractions and the overall mass conservation equation. Here, $X_{\beta}$ denotes the mole fraction of species $\beta$.

The subgrid scale (SGS) stress tensor, $T_{ij} = \overline{\rho u_i u_j} - \overline{\rho} \widetilde{u}_i \widetilde{u}_j$, is modeled within the framework of the Boussinesq hypothesis. The deviatoric component of this tensor is expressed as:
\begin{equation*}
T_{ij} - \frac{1}{3}T_{kk}\delta_{ij} = -2\mu_\sgs\left(\widetilde{S}_{ij} - \frac{1}{3}\widetilde{S}_{kk}\delta_{ij}\right) \,,
\end{equation*}
where $\mu_\sgs = \overline{\rho} \nu_\sgs$ is the SGS eddy viscosity, and $T_{kk}$ represents the isotropic part. The SGS mass flux closure employs a standard turbulent diffusivity assumption:
\begin{equation*}
T_{\varphi,i} = \overline{\rho \varphi u_i} - \overline{\rho} \widetilde{\varphi} \widetilde{u}_i = - \overline{\rho} D_\sgs \partial_i \widetilde{\varphi} \,,
\end{equation*}
where $\varphi$ is a generic scalar quantity, $D_\sgs = \nu_\sgs / \mathrm{Sc}_\sgs$ is the turbulent diffusivity, and $\mathrm{Sc}_\sgs$ is the turbulent Schmidt number. In the present study, the SGS eddy viscosity, $\mu_\sgs$, is computed using the WALE (Wall-Adapting Local Eddy-viscosity) model:
\begin{equation*}
\mu_\sgs = \overline{\rho}(C_w \Delta)^2\frac{\left(S_{ij}^dS_{ij}^d\right)^{3/2}}{\left(\widetilde{S}_{ij}\widetilde{S}_{ij}\right)^{5/2}+\left(S_{ij}^dS_{ij}^d\right)^{5/4}} \,,
\end{equation*}
where $\Delta = (\Delta x_1 \Delta x_2 \Delta x_3)^{1/3}$ is the characteristic grid scale, $C_w = C_s \sqrt{10.6}$ is the constant of the WALE model, and is the traceless tensor that writes
\begin{equation*}
S_{ij}^d=\frac12\left(\partial_k\widetilde{u}_i\partial_j\widetilde{u}_k+\partial_k\widetilde{u}_j\partial_i\widetilde{u}_k\right)-\frac13\partial_l\widetilde{u}_m\partial_m\widetilde{u}_l\delta_{ij}
\end{equation*}
The conservative form of the energy equation utilized herein is derived from the transport equation of the computable energy, $\overline{\rho}\breve{e} = \frac{\overline{p}}{\gamma-1} + \frac{\overline{\rho} \widetilde{u}_i \widetilde{u}_i}{2}$, as formulated by \citet{lee1993large}. Here, the breve symbol ($\breve{ }$) indicates a quantity based on primitive filtered variables and does not denote a filtering operation, unlike the overbar and tilde symbols. Thus, $\breve{e}$ represents the resolved total energy, which differs from the filtered total energy. Following \citet{vreman_priori_1995}, the transport equation for the computable energy is expressed as:
\begin{equation}\label{eq-etot_filtre_sci}
\begin{cases}
\partial_t\left(\overline{\rho} \breve{e}\right)
+\partial_j\left(\overline{\rho} \widetilde{u}_j \breve{e}\right)=
-\partial_j\left(\overline{p}\widetilde{u}_j\right)
+\partial_j\left(\widetilde{u}_i\overline{\tau}_{ij}\right)
-\partial_j\breve{q_j} + \mathcal{R}_e \,, \\
\mathcal{R}_e = - (B_1+B_2+B_3)+(B_4+B_5+B_6)-B_7 \,.
\end{cases}
\end{equation}
The $j$-th component of the computable molecular heat flux vector, $\breve{q_j}$, is approximated using the Hirschfelder and Curtiss formulation \cite{hirschfelder_molecular_1964}:
\begin{equation}
\label{eq-heat-flux_sci}
\breve{q_j} =
-\lambda(\widetilde T) \partial_j\widetilde{T}
+\sum_{\alpha=1}^N \overline{\rho} \widetilde{Y}_{\alpha} \widetilde{V}_{\alpha j} \widetilde{h}_{\alpha} \,,
\end{equation}
where $\lambda$ is the thermal conductivity of the multicomponent mixture, evaluated based on the filtered composition and temperature, and $\widetilde{h}_{\alpha}$ is the filtered enthalpy of species $\alpha$.
The subgrid terms $B_1$ to $B_7$ are defined as: $B_1=\partial_j(\overline{eu_j}-\overline{e}\widetilde{u}_j)$, $B_2=\overline{p\partial_j u_j}-\overline{p}\partial_j\widetilde{u_j}$, $B_3=\partial_j\left(T_{ij}\widetilde{u}_i\right)$, $B_4=T_{ij}\partial_j\widetilde{u}_i$, $B_5=\overline{\tau_{ij}\partial_j u_i}-\overline{\tau}_{ij}\partial_j\widetilde{u}_i$, $B_6=\partial_j\left(\overline{\tau_{ij} u_j}-\overline{\tau}_{ij}\widetilde{u}_i\right)$, and $B_7=\partial_j\left(\overline{q}_j-\breve{q}_j\right)$. Based on the numerical findings of \citet{vreman_priori_1995} and \citet{martin_subgridscale_2000}, the terms $B_4$, $B_6$, and $B_7$ are neglected due to their comparatively small contributions.

Applying the filtering operation to the transport equation of a passive scalar, $\xi$, yields:
\begin{equation}
\partial_t\left(\overline\rho \widetilde\xi\right)
+\partial_i\left(\overline\rho \widetilde u_i \widetilde\xi\right) =
-\partial_i\left(\breve{J}_{\xi,i}+J_{\xi,i}^\sgs\right)
+\partial_i\left(\overline J_{\xi,i}-\breve{J}_{\xi,i}\right) \,,
\label{eqzm_sci}
\end{equation}
where $\breve J_{\xi,i} = -\overline\rho \widetilde D \partial_i{\widetilde\xi}$ is the resolved molecular mass flux, $J_{\xi,i}^\sgs = \overline{\rho\xi u_i} - \overline\rho\widetilde\xi \widetilde u_i$ is the SGS mass flux, and $\left( \overline J_{\xi,i}-\breve J_{\xi,i} \right)$ represents the SGS molecular mass flux of the passive scalar, $\xi$. In this work, the subgrid contribution of the mass flux, $\left( \overline J_{\xi,i}-\breve J_{\xi,i} \right)$, is neglected, despite the apparent lack of prior investigations into its significance. The rationale for this simplification is that this term corresponds to molecular diffusion occurring at the subgrid level, which is assumed to be negligible in regimes of sufficiently developed turbulence \cite{martin_subgridscale_2000}.
\subsection{Numerical methods}
Numerical simulations are conducted using the in-house solver \textit{Izem}, which solves the three-dimensional, unsteady, compressible, reactive, multi-species Navier--Stokes equations. Inviscid fluxes are computed using a solution-adaptive finite-difference framework that combines an eighth-order central differencing scheme—suitable for resolving broadband turbulence—with a seventh-order Weighted Essentially Non-Oscillatory (WENO) scheme and Roe flux splitting in regions containing shocks. Shock detection is facilitated by an enhanced sensor based on normalized pressure and density gradients \cite{adams_high-resolution_1996}.
To minimize numerical dissipation in LES, a modified version of Ducros sensor \cite{ducros_large-eddy_1999} is employed, which is defined as
\begin{equation}
\varPhi_i = \frac{\left|-\widetilde{p}_{i-2} + 16\widetilde{p}_{i-1} - 30\widetilde{p}_i + 16\widetilde{p}_{i+1} - \widetilde{p}_{i+2}\right|}{\left|\widetilde{p}_{i-2} + 16\widetilde{p}_{i-1} + 30\widetilde{p}_i + 16\widetilde{p}_{i+1} + \widetilde{p}_{i+2}\right|} \cdot \frac{\left(\nabla \cdot \boldsymbol{\widetilde{u}}\right)^2}{\left(\nabla \cdot \boldsymbol{\widetilde{u}}\right)^2 + \left|\nabla \times \boldsymbol{\widetilde{u}}\right|^2},
\end{equation}
where $\varPhi_i$ is subsequently refined by considering its maximum over three neighboring cells $\varPhi_i = \max\left(\varPhi_{i+m}\right)$ for $m = -1, 0, 1$. This sensor is used to suppress subgrid-scale viscosity in regions where the velocity divergence exceeds a threshold of $0.7$.
Viscous and diffusive fluxes are discretized using an eighth-order central scheme, while time integration is performed using a third-order Total Variation Diminishing (TVD) Runge--Kutta method. Thermodynamic properties, including species enthalpies and specific heat capacities, are represented as seventh-degree polynomial functions of temperature, with coefficients obtained from the JANAF thermochemical tables.
The LES methodology, incorporating the aforementioned high-order numerical schemes and subgrid-scale modeling strategies, has been rigorously validated against experimental benchmarks, including compressible homogeneous isotropic turbulence \cite{boukharfane2021triple} and mixing layers \cite{boukharfane2021direct,boukharfane2021skewness}, underexpanded sonic jets issuing from nozzles \cite{baaziz2024large}, and transverse air jets interacting with a supersonic crossflow \cite{boukharfane2022reacting}.
\subsection{Numerical Setup}
In this study, the compressible Navier–Stokes equations are solved to investigate the mixing characteristics of various inert jet fuels injected into a supersonic crossflow. The freestream conditions correspond to a Mach number of 2.0, a static pressure of 56~kPa, and a static temperature of 1108~K. The Reynolds number, based on the injector diameter, is equal to 10,343. Because the state of the incoming boundary layer strongly influences the downstream development of the jet in crossflow, realistic inflow conditions are prescribed using the methodology proposed by Kawai and Lele~\cite{kawai_large_2008}. Specifically, steady profiles of the longitudinal velocity \( \tilde{u} \) and temperature \( \tilde{T} \) are imposed at the inlet. These profiles are extracted from a preliminary RANS simulation based on the \(k\text{-}\omega\) turbulence model and performed on the full-scale geometry of the {LAPCAT-II} test bench, which includes the nozzle, diffuser, combustion chamber, and divergent section. In this preliminary configuration, the walls are assumed isothermal at a temperature of 1100~K, and fuel injection is deactivated. Since the LES configuration considered here assumes adiabatic walls, the temperature profile is linearly extrapolated from the inflection point in order to ensure a smooth transition and physically consistent boundary layer development. The jets are injected perpendicularly into the supersonic crossflow at sonic conditions, i.e., the jet Mach number at the orifice exit is equal to 1.0. The velocity profile is prescribed using an error function formulation, which captures the presence of boundary layers at the injection interface.
A total of ten different injectant gases are considered in this study, including both pure species and various \ce{NH3}/\ce{H2}/\ce{N2} mixtures representative of alternative fuel blends relevant for high-speed propulsion applications. All injectants are injected at the same total pressure and temperature, namely \( p_{t,j} = 958{,}055~\text{Pa} \) and \( T_{t,j} = 300~\text{K} \), in order to isolate the effects of their thermodynamic and transport properties on the jet development. Table~\ref{tab:injection_conditions} summarizes the injection conditions for all fuels considered, including their composition, mean molar mass, specific heat ratio \( \gamma_j \), as well as key non-dimensional parameters such as the velocity ratio \( U_j/U_\infty \), the density ratio \( \rho_j/\rho_\infty \), the static pressure ratio \( p_j/p_\infty \), and the resulting mass flow rate \( \dot{m}_j \).

\begin{table}[h]
\centering
\caption{Summary of injection conditions ($D = 2.0 \, \text{mm}$, $M_j = 1.0$, $M_\infty = 2.0$)}
\label{tab:injection_conditions}
\resizebox{\textwidth}{!}{%
\begin{tabular}{lcccccccc}
\toprule
\textbf{Injectant gas} & Composition (\% by vol.) & $J$ & Mean molar mass (g/mol) & $\gamma_j$ & $U_j / U_\infty$ & $\rho_j / \rho_\infty$ & $p_j / p_\infty$ & $\dot{m}_j$ (g/s) \\
\midrule
\ce{H2}                           & 100                   & 2.32 & 2    & 1.41 & 0.91 & 2.75 & 17.08 & 1.85 \\
\ce{NH3}                          & 100                   & 2.16 & 17   & 1.31 & 0.30 & 23.30 & 16.55 & 5.20 \\
\ce{CH4}                          & 100                   & 2.16 & 16   & 1.32 & 0.31 & 21.95 & 16.57 & 5.06 \\
\ce{C2H4}                         & 100                   & 2.09 & 28   & 1.27 & 0.23 & 38.38 & 16.31 & 6.57 \\
\ce{CO}/\ce{H2}                   & $30/70$               & 2.32 & 2.8  & 1.41 & 0.77 & 3.82 & 17.08 & 2.18 \\
\ce{NH3}/\ce{H2}/\ce{N2}/\ce{O2} (S0) & $20/25/45/10$         & 2.30 & 6.41 & 1.40 & 0.51 & 8.78 & 17.84 & 3.30 \\
\ce{NH3}/\ce{H2}/\ce{N2} (S1)     & $80/15/5$             & 2.30 & 8.12 & 1.37 & 0.44 & 11.10 & 16.84 & 3.66 \\
\ce{NH3}/\ce{H2}/\ce{N2} (S2)     & $60/30/10$            & 2.28 & 5.33 & 1.39 & 0.55 & 7.29  & 16.95 & 2.99 \\
\ce{NH3}/\ce{H2}/\ce{N2} (S3)     & $40/45/15$            & 2.30 & 3.96 & 1.40 & 0.65 & 5.42  & 17.02 & 2.59 \\
\ce{NH3}/\ce{H2}/\ce{N2} (S4)     & $20/60/20$            & 2.31 & 3.15 & 1.41 & 0.73 & 4.32  & 17.05 & 2.32 \\
\bottomrule
\end{tabular}}
\end{table}

Non-slip and adiabatic wall boundary conditions are imposed in the vicinity of the fuel injection region on the lower boundary of the computational domain. At the upper, rear, and front boundaries, extrapolation conditions are employed in conjunction with sponge layers and mesh coarsening to suppress the formation and reflection of non-physical numerical waves. The flow field is initialized throughout the domain with supersonic air inflow conditions.
\subsection{Computational Setup and Grid Resolution}
\begin{figure}[ht!]
\centering
\includegraphics[width=0.99\textwidth]{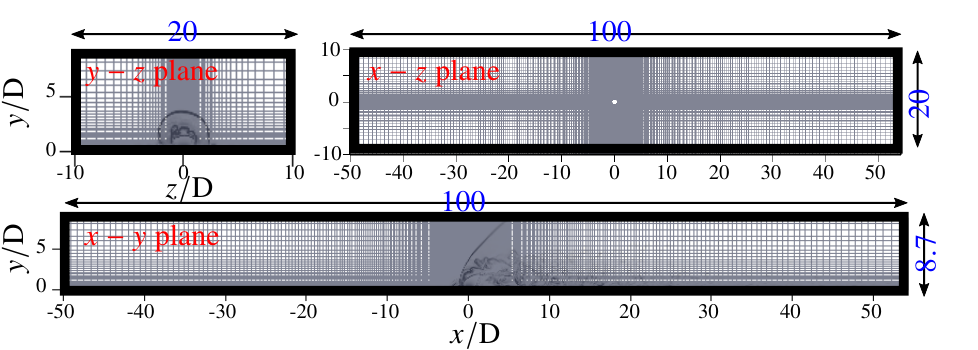}
\caption{Representation of the computational domain and the mesh}
\label{fig:mesh-domaine}
\end{figure}

\begin{figure}[ht!]
\centering
\includegraphics[width=0.69\textwidth]{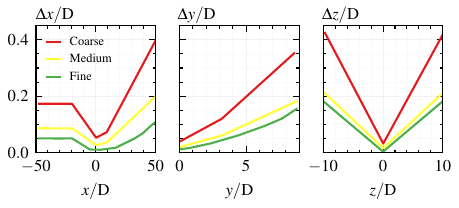}
\caption{Grid spacing normalized by the jet diameter $\mathrm{D}$ along the three spatial directions for the coarse (red), medium (yellow), and fine (green) meshes. From left to right: streamwise spacing $\Delta x / D$, vertical spacing $\Delta y / D$, and lateral spacing $\Delta z / D$.}
\label{fig:mesh-comparison}
\end{figure}

\begin{figure}[ht!]
\centering
\subfloat[]{\includegraphics[width=0.79\textwidth]{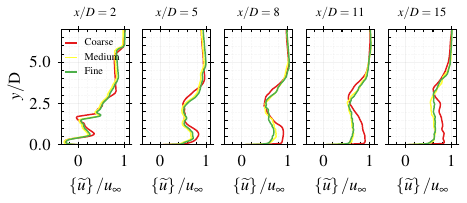}}\\
\subfloat[]{\includegraphics[width=0.79\textwidth]{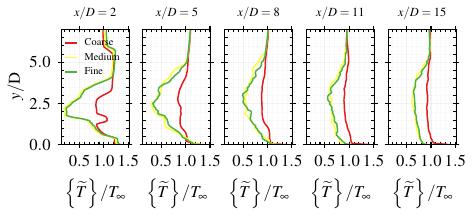}}
\caption{Mesh sensitivity validation: (a) Normalized mean streamwise velocity and (b) temperature at different streamwise locations of the $z/\mathrm{D}=0$ slice.}
\label{fig:compTU_meshes}
\end{figure}

The computational domain and grid are illustrated in Fig.~\ref{fig:mesh-domaine}. The jet are located on the bottom wall of the computational domain and consists of a circular orifice with a diameter of 2~mm, positioned such that the center of the orifice coincides with the origin of the coordinate system. The jet is centered centered $50\mathrm{D}$ downstream from the inlet along the centerline. The main computational domain extends $100\mathrm{D}$ in the streamwise direction, $8.75\mathrm{D}$ in the wall-normal direction, and $20\mathrm{D}$ in the spanwise direction.
To mitigate turbulent fluctuations and suppress non-physical reflections at the boundaries, buffer zones with gradually coarsened mesh are implemented outside the main domain. These buffer regions are excluded from the analysis.
A mesh independence study is conducted using three non-uniform Cartesian grids of increasing resolution, comprising approximately 9.6 million, 60.8 million, and 169.7 million cells, respectively. The distribution of cell sizes in the streamwise, vertical, and lateral directions is shown in Fig.~\ref{fig:mesh-comparison}. The computed time elapse to at approximately \( t = 150D/u_\infty = 230 \, \mu s \). Starting from \( t = 110D/u_\infty = 168 \, \mu s \), fields are recorded with an acquisition frequency of $8.70$ MHz, corresponding to every $0.115$ $\mu$s, thus allowing the collection of more than $530$ fields, which allows to obtain statistically meaningful turbulence properties in the temporal average operation. The average time step of the calculation (CFL limiting) is on the order of $1.5$ ns.
All computational grids employ local refinement near the injection center, with a gradual coarsening of resolution toward the domain boundaries. The finest mesh density is achieved at the core region around $x/\mathrm{D} = 0$, $y/\mathrm{D} = 0$, and $z/D = 0$. The mean and normalized velocity and temperature profiles along the wall-normal direction at the mid-plane intersecting the jet centerline ($z/D = 0$), evaluated at various downstream streamwise positions for the \ce{NH3} injectant fuel are shown in Fig.~\ref{fig:compTU_meshes}. It is evident that the velocity and temperature distributions obtained from the moderate and refined meshes are in close agreement, indicating good resolution of the jet dynamics. In contrast, the coarse mesh exhibits significant deviations, especially in the near-field region just downstream of the injection point. Based on this comparative analysis, the moderate mesh is deemed both sufficient and computationally efficient for capturing the essential flow features in the present study.
\section{Results and Discussion}
\subsection{Instantaneous Structure}
\begin{figure}[ht!]
\centering
\includegraphics[width=0.99\textwidth]{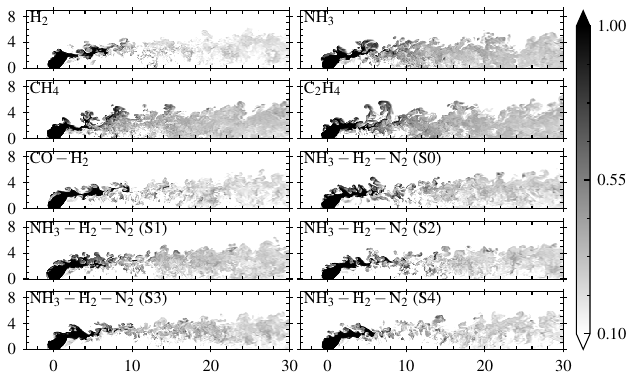}
\caption{Instantaneous numerical schlieren-like visualization using different type of fuels}
\label{fig:schlierenInst}
\end{figure}

%
An instantaneous visualization of the mixture fraction flowfields is shown in Fig.~\ref{fig:schlierenInst}. It can be seen that for all the cases the shock interacts with the upstream boundary layer, which will separate to form the $\lambda$-shock structure. It is also possible to identify the mixing layer developing above the jet, via Kelvin-Helmholtz type instabilities, thus allowing the evolution of the jet penetration height into the flow to be seen; this latter characteristic strongly depends on the parameter $J$. Indeed, the larger the value of $J$, the greater the penetration height, as is the interaction between the jet, the main flow, the boundary layer, and the intensity of the curved shock; conversely, the deflection and deformation of the barrel shock is more significant for low values of $J$ \citep{gamba_ignition_2015}.
In all the cases examined, fuel plumes that have been injected exhibit characteristic signatures of JICF phenomena. Included among these are an observable jet penetration into the supersonic stream, the formation of Kelvin-Helmholtz-like instabilities along the shear layer of the jet, manifested as rotating vortical structures (darker areas representing regions of high density gradient), the creation of a bow shock wave in front of the jet through the deflection of the supersonic stream, and a turbulent wake region of complex nature downstream. Nonetheless, the degree and shape of these characteristics are greatly affected by the thermophysical properties and injection conditions of the various fuels. For comparison between the pure fuels, hydrogen (H2) possesses the deepest penetration, which can most likely be explained by its high momentum flux ratio ($J = 2.32$) and high jet velocity relative to the crossflow ($U_j / U_\infty = 0.91$). Complex turbulent structures at a small scale are displayed by the shear layer in the hydrogen jet. Conversely, bigger hydrocarbons such as ammonia (\ce{NH3}), methane (\ce{CH4}), and ethylene (\ce{C2H4}) are characterized by a progressive decrease in penetration accompanied by a greater curvature of the trajectory of the jet. This is attributed to their lower momentum flux ratios and far smaller jet velocity ratios, alongside high density ratios ($\rho_j / \rho_\infty$). The vortical structures obtained in schlieren photographs for these fuels also seem to be more intense and extensive. The CO/H2 blend, with its high hydrogen content, has an intermediate penetration depth between pure hydrogen and the heavier fuels, as might be anticipated from its intermediate injection conditions. For the \ce{NH3}/\ce{H2}/\ce{N2} blends (S0-S4), a trend is evident: increasing the hydrogen fraction in the injectant has the effect of increasing jet penetration. The S1 blend, which contains the highest ammonia fraction, has the lowest penetration of this group and is comparable to pure ammonia. In contrast, the S4 blend, being rich in hydrogen, results in a penetration depth that approaches the one observed in the pure \ce{CO}/\ce{H2} experiment. The intermediate blends (S0, S2, and S3) all exhibit penetration depths that correspond to their respective hydrogen content. The oxygen content of the S0 blend is not seen to have any apparent effect on the jet penetration compared to the other blends of similar hydrogen content. The schlieren images collectively demonstrate the wide-ranging influence of fuel composition and the corresponding injection parameters in controlling the macroscopic behavior and characteristics of the density field of sonic jets injected transversely into a supersonic crossflow.

\begin{figure}[ht!]
\centering
\includegraphics[width=0.99\textwidth]{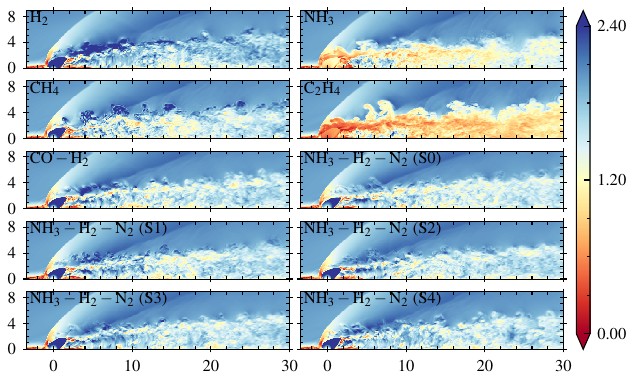}
\caption{Instantaneous visualization of the Mach number using different type of fuels}
\label{fig:machInst}
\end{figure}

The Mach number fields depicted in Fig.~\ref{fig:machInst} provide crucial insights into the complex interaction between the sonic fuel jets and the Mach 2 supersonic crossflow. A consistent feature across all cases is the presence of the undisturbed supersonic crossflow, characterized by a Mach number of approximately 2.0 in the upper regions of the computational domain. The injection of the sonic jets introduces significant perturbations to this uniform flow, generating regions of both reduced and elevated Mach numbers in the vicinity of the jet and its downstream wake.
A prominent bow shock wave is observed upstream of each jet, signifying the deceleration of the supersonic crossflow as it encounters the obstruction posed by the injected fluid. The intensity and standoff distance of this bow shock vary considerably with the fuel type. For pure fuels, hydrogen (\ce{H2}) generates a particularly strong bow shock with a noticeable upstream influence, while heavier hydrocarbons like ammonia (\ce{NH3}), methane (\ce{CH4}), and especially ethylene (\ce{C2H4}) exhibit progressively weaker bow shocks, indicating a less forceful interaction with the freestream. The core of the injected jets generally displays lower Mach numbers compared to the crossflow, particularly near the injection orifice, due to expansion and mixing with slower-moving near-wall fluid. Downstream, expansion waves can lead to localized increases in Mach number. The wake region is characterized by a complex Mach number distribution resulting from turbulence, flow separation, and vortical structures.
For the fuel mixtures, the \ce{CO}/\ce{H2} blend shows an intermediate bow shock strength and jet core Mach number retention, consistent with its composition. The \ce{NH3}/\ce{H2}/\ce{N2} mixtures (S0-S4) reveal a clear correlation between hydrogen content and the Mach number field characteristics. Mixtures with higher ammonia fractions (like S1) exhibit weaker bow shocks and rapid deceleration of the jet core, while those richer in hydrogen (like S4) produce more pronounced bow shocks and maintain higher Mach numbers within the jet further downstream, similar to the \ce{CO}/\ce{H2} case. The Mach number fields underscore the significant role of the injected fuel's properties in dictating the nature and extent of its interaction with the supersonic crossflow, influencing the shock wave structure, jet penetration, and the downstream flow dynamics.

\begin{figure}[ht!]
\centering
\subfloat[]{\includegraphics[width=0.69\textwidth]{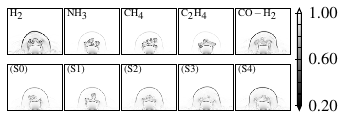}}\\
\subfloat[]{\includegraphics[width=0.69\textwidth]{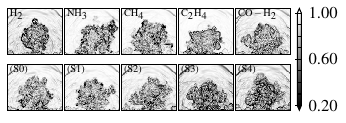}}\\
\subfloat[]{\includegraphics[width=0.69\textwidth]{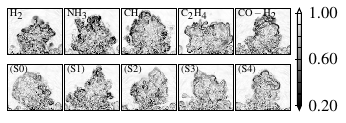}}
\caption{Numerical schlieren at three downstream locations ((a) $0\mathrm{D}$, (b) $25\mathrm{D}$, and (c) $35\mathrm{D}$) downstream of injection for different fuels.}
\label{fig:instYZ_p_sDensity}
\end{figure}

The numerical Schlieren $S_{\bar\rho}=\exp\left(-\alpha \frac{\|\nabla\rho\|}{\max \|\nabla\rho\|}\right)$ based on the density at three distinct downstream locations from the injection point ($0\mathrm{D}$, $25\mathrm{D}$, and $35\mathrm{D}$) are shown in Fig.~\ref{fig:instYZ_p_sDensity}. 
At the injection plane ($0\mathrm{D}$), a detached bow shock is observed upstream for all fuels, with the jet plume exhibiting limited initial penetration and a developing shear layer. The characteristics at this stage are strongly influenced by fuel-dependent injection conditions such as momentum flux ratio ($J$) and density ratio ($\rho_j / \rho_\infty$).
Moving downstream to $25\mathrm{D}$, the jet undergoes significant changes, with coherent structures breaking down into turbulence, leading to enhanced mixing and increased vertical penetration. The bow shock becomes less prominent as the interaction spreads. Fuels with higher initial momentum tend to show greater penetration.
Further downstream at $35\mathrm{D}$, turbulent mixing dominates, resulting in a larger mixed region with finer-scale structures and significant lateral spreading, likely due to the counter-rotating vortex pair. The vertical penetration might approach its maximum, and any trace of the bow shock disappears. The influence of initial fuel properties on penetration might become less distinct as mixing becomes the primary driver of the flow field.
Comparing the behavior across different fuels reveals that hydrogen-rich injectants (pure \ce{H2} and hydrogen-rich blends) generally exhibit greater initial and downstream penetration, along with more extensive mixing. Heavier hydrocarbons (e.g., \ce{C2H4}) tend to have lower initial penetration but achieve significant mixing further downstream. Fuels like \ce{NH3} and \ce{CH4} show intermediate characteristics, while the \ce{CO}/\ce{H2} blend behaves similarly to hydrogen-rich fuels.
The evolution from the near-field to the far-field demonstrates a transition from a flow dominated by initial momentum and blockage effects to one governed by turbulent mixing and large-scale vortical structures. While the initial fuel properties are crucial in setting the stage for the jet's development, the turbulent mixing processes become increasingly important in determining the jet's ultimate penetration, spread, and interaction with the supersonic crossflow at greater distances from the injection.
%
\subsection{Time-Averaged Averaged Field}
\begin{figure}[ht!]
\centering
\subfloat[]{\includegraphics[width=0.69\textwidth]{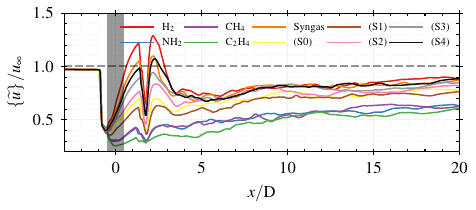}}\\
\subfloat[]{\includegraphics[width=0.69\textwidth]{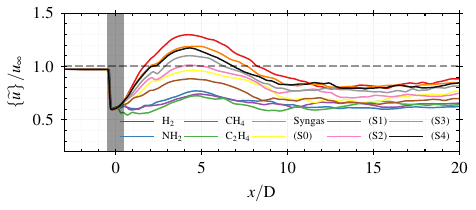}}
\caption{Normalized streamwise velocity ($\{\tilde{u}\} / u_\infty$) at (a) ($y/\mathrm{D} = 2$, $z/D=0$) and (b) ($y/\mathrm{D} = 3$, $z/\mathrm{D}=0$) for different fuels.}
\label{fig:normalized_velocity_yd23}
\end{figure}

%
The normalized streamwise velocity profiles at two different heights above the injection wall $y/\mathrm{D} = 2$ and $y/\mathrm{D} = 3$ in the lateral slice passing through the center inejctor is diplayed in Fig.~\ref{fig:normalized_velocity_yd23}.
At both locations, a decrease in the normalized streamwise velocity is observed near the injection ($x/\mathrm{D} \approx 0$) for all fuels, signifying the jet plume's blockage effect. However, the magnitude of this initial velocity drop is generally smaller at $y/\mathrm{D} = 3$ compared to $y/\mathrm{D} = 2$, indicating a reduced direct blockage effect of the jet as we move away from the wall. The upstream influence of the jet, likely associated with the detached bow shock, is evident in the velocity drop starting slightly upstream of the injection at both heights.
Downstream of this initial deficit, several fuels exhibit a velocity peak exceeding the freestream value, suggesting localized acceleration or compression due to the complex jet-crossflow interaction. This peak is present at both $y/\mathrm{D} = 2$ and $y/\mathrm{D} = 3$, but its magnitude and location might shift. In general, the peaks appear to be more pronounced and occur slightly further downstream at $y/\mathrm{D} = 3$ compared to $y/\mathrm{D} = 2$, which could be related to the vertical evolution of the bow shock and counter-rotating vortex pair. The velocity eventually recovers towards the freestream value at both heights, but the recovery rate and asymptotic values might differ. The recovery at $y/\mathrm{D} = 3$ could be faster due to the reduced influence of the near-wall wake.
Comparing the fuel dependence, the relative ordering and separation of the velocity curves are broadly similar at both heights. However, there are some nuances in the magnitudes and locations of the key features. For instance, fuels like \ce{H2} and Syngas tend to maintain higher velocities downstream compared to \ce{NH3} and \ce{C2H4} at both locations. The blended fuels show intermediate behaviors depending on their composition at both heights. The differences in velocity profiles between fuels indicate that the injectant's properties continue to influence the jet's interaction with the crossflow at both elevations.

\section{Analysis of Jet Penetration Trajectories}

Several definitions of jet trajectory have been proposed in the literature, such as the locus of local velocity maxima~\cite{kamotani1972experiments} or the locus of local scalar concentration maxima~\cite{ben-yakar_time_2006}. However, both velocity and scalar fields often exhibit multiple local maxima~\cite{yuan1998trajectory}, complicating the automatic determination of the jet trajectory. Following the approach of \citet{muppidi2005study}, the present study defines the jet trajectory as the streamline emanating from the center of the jet exit (referred to as the center streamline) on the time-averaged symmetry plane. This definition provides a more accurate representation of the path followed by the jet fluid and is particularly well suited for numerical simulations.
\begin{figure}[ht!]
\centering
\includegraphics[width=0.69\textwidth]{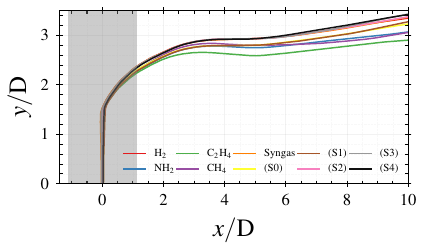}
\caption{Comparison of jet penetration, with both streamwise and transverse coordinates normalized by the jet diameter ($\mathrm{D}$), for different types of fuel.}
\label{fig:jetpen1}
\end{figure}
Figure~\ref{fig:jetpen1} shows the time-averaged trajectory of the jet centerline for various fuels injected transversely into a supersonic crossflow, illustrating the normalized vertical penetration ($y/\mathrm{D}$) as a function of the normalized streamwise distance ($x/\mathrm{D}$). Across all tested fuels, a common feature is observed: a rapid initial increase in penetration height near the injection orifice ($x/\mathrm{D} = 0$), driven by the jet’s transverse momentum. As the jets progress downstream, the supersonic crossflow exerts an increasing influence, causing the trajectories to bend in the streamwise direction. The extent and rate of this bending vary significantly depending on the fuel properties.
At larger downstream distances, the penetration height tends to either plateau or increase more gradually, suggesting a balance between the jet's residual transverse momentum and the deflecting force of the crossflow. Each fuel exhibits a distinct penetration trajectory, highlighting the strong influence of fuel composition and associated injection parameters.
Hydrogen (\ce{H2}) demonstrates the highest overall penetration, maintaining a steep trajectory even at larger $x/\mathrm{D}$, indicative of its strong resistance to the crossflow’s bending effect, consistent with its high momentum flux ratio ($J$) and jet velocity ratio ($u_j / u_\infty$). Syngas (\ce{CO}/\ce{H2}) follows closely, owing to its high hydrogen content and comparable $J$. The \ce{NH3}/\ce{H2}/\ce{N2} mixtures (S3 and S4) show intermediate penetration levels, with a clear trend: higher hydrogen content (as in S4) results in greater penetration.
Methane (\ce{CH4}) and the S2 mixture exhibit moderate penetration with more pronounced bending compared to hydrogen-rich fuels. The S0 and S1 mixtures show lower penetration levels, with S1 (high ammonia content) displaying the shallowest trajectory among the S-series. Ethylene (\ce{C2H4}) exhibits the lowest penetration, characterized by rapid bending and a low plateau, consistent with its low momentum flux and velocity ratios and high density ratio.
Overall, the observed penetration trajectories correlate strongly with the momentum flux ratio $J$: higher $J$ leads to greater penetration. The jet velocity ratio ($u_j / u_\infty$) also plays a significant role, as evidenced by hydrogen's superior performance. Conversely, a higher density ratio ($\rho_j / \rho_\infty$) tends to reduce penetration, as seen for ethylene. The \ce{NH3}/\ce{H2}/\ce{N2} mixtures underscore the role of fuel composition, where varying proportions of light and heavy components directly impact the jet behavior. 

\begin{figure}[ht!]
\centering
\includegraphics[width=0.69\textwidth]{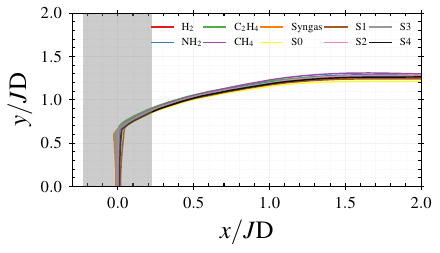}
\caption{Comparison of jet penetration, with both streamwise and transverse coordinates normalized by the product of the momentum flux ratio ($J$) and jet diameter ($\mathrm{D}$), for different types of fuel.}
\label{fig:jetpen2}
\end{figure}
Figure~\ref{fig:jetpen2} illustrates the trajectories of the jet centerline averaged over time, with both the streamwise and transverse coordinates normalized by the product of the momentum flux ratio ($J$) and the jet diameter ($\mathrm{D}$). This approach to normalization provides an alternative viewpoint in contrast to merely scaling by $\mathrm{D}$, emphasizing the significant influence of $J$ in determining the initial behavior of the jet.
A striking observation is the strong reduction of penetration trajectories in the near-field region ($x/J\mathrm{D} \lesssim 1$) for all fuels. This universality strongly suggests that the momentum flux ratio serves as the governing scaling parameter for the early penetration of transverse jets into a supersonic crossflow, thus effectively unifying the early interaction dynamics for different fuels.
As the normalized streamwise distance ($x/J\mathrm{D}$) increases, the distinction between the trajectories is apparent. This divergence in the far field shows that while $J$ dictates the initial scaling, it fails to capture the intricate physical processes that determine the downstream behavior of the jet. Variables such as thermophysical properties (e.g., density ratio and specific heat ratio), their influence on mixing efficiency, the generation of large-scale turbulent structures, and their interactions with the shock system take on more importance at larger distances.
Therefore, although normalization by the use of parameter $J$ affords a basic foundation for the understanding of the initial scaling of jet penetration, the persistence of divergence downstream serves to underscore the intrinsic complexity inherent in the jet-in-crossflow phenomenon. Near-field collapse suggests that for predictions of initial penetration as well as idealized models, $J$ is a significant variable. Yet, precise prediction of mixing and combustion dynamics downstream necessitates a thorough consideration of fuel-specific properties and their effects on turbulence.
Ultimately, small discrepancies found in the near-field can be ascribed to second-order effects or to random variations in other dimensionless parameters not accounted for in the $J$-based scaling approach. Examination of these discrepancies may provide further insight into complicated behaviors of various fuels under the same momentum flux conditions.

\section{Conclusion}

In this study, high-fidelity LES has been utilized to investigate the influence of fuel type on the flow field and mixing characteristics of a transverse sonic jet injected into a Mach 2 supersonic crossflow. By maintaining an approximately constant momentum flux ratio across six fuels—hydrogen, methane, ethylene, ammonia, syngas, and a synthetic blend—we have been able to decouple the influence of their respective thermophysical properties on the windward mixing layer and the overall development of the jet.
The real-time visualizations of the flow field, through numerical schlieren and Mach number contours, exhibited significant differences in the bow shock structure, jet penetration, and development of large-scale vortical structures that were a function of the fuel type. Lighter fuels with higher momentum and velocity ratios, such as hydrogen and syngas, exhibited greater initial and total penetration into the supersonic crossflow, as well as finer-scale turbulent structures within their shear layers. On the other hand, heavier hydrocarbons such as ethylene exhibited less penetration with more rapid deflection of the jet. The \ce{NH3}/\ce{H2}/\ce{N2} mixtures exhibited intermediate behavior directly proportional to their hydrogen concentration.
Comparison of the time-averaged jet centerline trajectories, normalized by the jet diameter, also offered quantification of these observations. Hydrogen and syngas penetrated the most, and ethylene penetrated the least. The \ce{NH3}/\ce{H2}/\ce{N2} mixtures exhibited a strongly pronounced trend of increasing penetration as the hydrogen content increased. It is noteworthy that the normalization of the streamwise and transverse coordinates by the jet diameter and momentum flux ratio resulted in an excellent convergence of the jets' trajectories in the near-field zone. This result demonstrates the overwhelming significance of the momentum flux ratio as an essential scaling parameter for the initial penetration dynamics, actually unifying the behavior of various fuels in the immediate surroundings of the injection. But the later divergence of these normalized trajectories in the far-field emphasized the limitations of $J$ as the single scaling parameter for the overall evolution of the jet, pointing to the increasing role of other fuel-specific thermophysical properties and their effects on downstream mixing and turbulence.
In summary, the present study has elucidated the fuel-dependent mixing characteristics in supersonic transverse jets. The findings indicate that although the momentum flux ratio primarily influences the initial jet penetration, the subsequent development and mixing characteristics are strongly dependent on the individual molecular weight, heat capacity ratio, and density of the injected fuel. The results of this study are important to the advancement and refinement of fuel injection methods in high-speed propulsion systems, especially with the growing focus on alternative and synthetic multicomponent fuels with a range of thermophysical properties. Further studies can address a more in-depth examination of the scalar mixing fields and turbulence statistics to provide a clearer picture of the impact of changes in fuel composition on the effectiveness and uniformity of fuel-air mixing under these complicated flow conditions.

\subsection*{Acknowledgement}
This work was supported by OCP Group (Morocco). The authors gratefully acknowledge the support and computing resources from the African Supercomputing Center (ASCC) at UM6P (Morocco).
%

\bibliographystyle{plainnat}
\bibliography{main.bib}
\end{document}